\documentclass[twocolumn, astrosymb]{aastex631}

\newcommand\quasar{SDSS J0252$-$0028}
\usepackage{amsmath}
\newcommand{\RN}[1]{%
  \textup{\uppercase\expandafter{\romannumeral#1}}%
}

\shorttitle{SDSS J0252-0028}
\shortauthors{Foord et al.}

\begin{document}

\title{Investigating the Accretion Nature of Binary Supermassive Black Hole Candidate SDSS J025214.67-002813.7}

\author[0000-0002-1616-1701]{Adi Foord}
\affiliation{Kavli Institute of Particle Astrophysics and Cosmology, Stanford University, Stanford, CA 94305, USA}

\author[0000-0003-0049-5210]{Xin Liu}
\affiliation{Department of Astronomy, University of Illinois at Urbana-Champaign, Urbana, IL 61801, USA}
\affiliation{National Center for Supercomputing Applications, University of Illinois at Urbana-Champaign, Urbana, IL 61801, USA}

\author[0000-0002-1146-0198]{Kayhan G\"{u}ltekin}
\affiliation{Department of Astronomy and Astrophysics, University of Michigan, Ann Arbor, MI 48109}

\author[0000-0001-9379-6519]{Kevin Whitley}
\affiliation{Department of Astronomy and Astrophysics, University of Michigan, Ann Arbor, MI 48109}

\author[0000-0003-3922-5007]{Fangzheng Shi}
\affiliation{School of Astronomy and Space Science, Nanjing University, Nanjing 210023, China}
\affiliation{Key Laboratory of Modern Astronomy and Astrophysics, Nanjing University, Nanjing 210023,
China}

\author[0000-0002-9932-1298]{Yu-Ching Chen}
\affiliation{Department of Astronomy, University of Illinois at Urbana-Champaign, Urbana, IL 61801, USA}
\affiliation{National Center for Supercomputing Applications, University of Illinois at Urbana-Champaign, Urbana, IL 61801, USA}
\affiliation{Center for AstroPhysical Surveys, National Center for Supercomputing Applications, Urbana, IL, 61801, USA}





\begin{abstract}
We present results on a multi-wavelength analysis of SDSS J025214.67-002813.7, a system which has been previously classified as a binary AGN candidate based on periodic signals detected in the optical light curves. We use available radio$-$X-ray observations of the system to investigate the true accretion nature. Analyzing new observations from \emph{XMM-Newton} and \emph{NuSTAR}, we characterize the X-ray emission and search for evidence of circumbinary accretion. Although the 0.5--10 keV spectrum shows evidence of an  additional  soft  emission  component, possibly due to extended emission from hot nuclear gas, we find the spectral shape consistent with a single AGN. Compiling a full multi-wavelength SED, we also search for signs of circumbinary accretion, such as a ``notch" in the continuum due to the presence of minidisks. We find that the radio--optical emission agrees with the SED of a standard, radio-quiet, AGN, however there is a large deficit in emission blueward of $\sim$1400 ${\rm\AA}$. Although this deficit in emission can plausibly be attributed to a binary AGN system, we find that the SED of \quasar{} is better explained by emission from a reddened, single AGN. However, future studies on the expected hard X-ray emission associated with binary AGN (especially in the unequal-mass regime), will allow for more rigorous analyses of the binary AGN hypothesis.
\end{abstract}



\section{Introduction} \label{sec:intro}
\par A binary supermassive black hole (SMBH) represents the final stage of a galaxy merger, where the two massive host galaxies have likely been interacting for hundreds of megayears to gigayears \citep{Begelman1980}. The merging system is classified as a binary when the SMBHs are gravitationally bound in a Keplerian orbit, and for a wide range SMBH masses and host galaxy environments this occurs at orbital separations $<$10 pc \citep{Mayer2007, Dotti2007, Khan2012}. The fate and final coalescence of the system strongly depends on the amount of matter the SMBHs can interact with \citep{Sesana2007, Merritt2007}. As the last stage before coalescence, binary SMBHs represent an observable link between galaxy mergers and gravitational wave events. They are strong emitters of low-frequency gravitational waves, which are expected to dominate the gravitational wave background signal detected by pulsar timing arrays (PTAs; \citealt{Burke-Spolaor2019}), and they are direct precursors to gravitational-wave events detectable by future space-based laser interferometers \citep{Sesana2007}.
\par Currently, detections of binary SMBHs are limited to systems where both black holes are actively accreting (active galactic nuclei, or AGN) and emitting light across the electromagnetic spectrum. Given the small physical separations between the two SMBHs, the angular resolution afforded by radio interferomerty is required to spatially resolve binary AGN systems, such as in the serendiptious discovery of 0402+379 \citep{Rodriguez2006}. However, radio detections of binary AGN remain limited to relatively low-redshifts (where separations below 1 pc can only be probed up to $z=0.1$) and systems with two radio-bright AGN (where only $\sim$15\% of AGN are expected to be radio loud; see, e.g., \citealt{Hooper1995, Kellermann2016}). On top of this, blind surveys with the VLBI networks are limited by the narrow ($\sim$ arcsec-scale) fields of view. As a result, the number of confirmed binary AGN remains small. 
\par Given the difficulty associated with directly detecting binary AGN, many indirect detection techniques are used to search for the elusive systems. One of the most popular methods is to identify quasars with photometric variability, and in particular, periodic signals in their light curves. Periodicity in light curves may arise from various reasons, such as jet precession (where the presence of a companion can introduce a periodicity in the velocity of an otherwise straight jet; see \citealt{Hardee1994, Deane2014}), the dynamics of a secondary periodically intercepting the primary SMBH’s accretion disk (i.e., OJ 287; see \citealt{Valtonen2008} and the recent review in \citealt{Dey2019}), or accretion via a circumbinary disk (\citealt{Hayasaki2007, MacFadyen2008, Roedig2012, Roedig2014, D'Orazio2013, Farris2014}). Recently, systematic scans of large areas of the sky carried out via time domain surveys allow for statistical searches for periodic variability in large samples of quasars \citep{Valtonen2008, Graham2015, Charisi2016, Li2016, Liu2019, Bon2016, Chen2020}. Such analyses are complicated by the fact that AGN light curves are characterized by stochastic, red noise variability; it has been shown that this red noise can be misidentified as a periodic signal with time baselines fewer than 5 periods \citep{Vaughan2016}. Although these surveys have collectively found over 100 binary AGN candidates, to date there are no confirmed binary AGN with separations at the sub-milliparsec scale.
\par A different approach is to use available multi-wavelength observations to search for evidence of circumbinary accretion (see, e.g., \citealt{Foord2017b}). Circumbinary accretion is expected at small separations (when the typical accretion disk size is larger than the binary separation $a$; \citealt{Milosavljevic2005}). Here, two accretion disks around each supermassive black hole (``mini-disks'') are surrounded by a larger, circumbinary accretion disk. The mini-disks extend to a tidal truncation radius (i.e., where the tidal torques of the disks balance the viscous torque of the circumbinary accretion disk), where values are less than $a$/2 (\citealt{Paczynski1977}, but see \citealt{Roedig2014} for how the radius depends on the mass ratio of the system). In later stages, the angular momentum of material falling from the circumbinary accretion disk may exceed that of the inner-most stable orbits (ISCO) of the SMBHs, and material will fall directly into each SMBH (see \citealt{Gultekin&Miller2012, Tanaka&Haiman2013, Gold2013, Roedig2014}). Each accretion scenario will manifest differently in the spectral energy distribution (SED) of a binary AGN, where dips in the optical -- UV bands or very little high-energy emission is observed (see \citealt{Roedig2014, Foord2017b}). Complications to these simple scenarios naturally arise when accounting for stream shocking, or when supersonic material from the circumbinary accretion disk hits the outer edge of the mini disks. The high-energy signal associated with these types of events can wash out any dip in the optical regime of the SED \citep{Roedig2014, Farris2015a, Farris2015b}. This is especially true for systems where the dips in emission are expected to be subtle with respect to the overall shape of the SED (which depend on the mass ratio between the secondary and primary SMBH and the accretion rates of each SMBH see Section 3 for more details).
\par Here we present a multi-wavelength analysis of binary AGN candidate SDSS J025214.67$-$002813.7 (hereafter \quasar{}), located at $z=1.53$. It has an estimated virial black hole mass of $M=10^{8.4\pm0.1}\ M_{\odot}$, physical separation on the order of $\approx$ 4 miliparsecs (or 200 Schwarzschild radii), and a binary mass ratio on the order of $\sim$0.1 \citep{Liao2021}. \quasar{} was part of a large systematic search for periodic light curves in 625 quasars via combining Dark Energy Survey Supernova (DES-SN) Y6 observations with archival SDSS-S82
data \citep{Chen2020}. Briefly, quasars were flagged as having significant periodicity in their light curves if (1) at least two photometric bands had a 3$\sigma$ detection of the same periodicity in the periodogram analysis, (2) the detected periodicity was the dominant signal with respect to the background (red noise) and (3) the same periodicity was also identified in the autocorrelation function (ACF) analysis (see \citealt{Liao2021} for more details). Among the 5 quasars flagged as significant periodic candidates, \quasar{} was the most significant detection based on $\sim$4.6 cycles detected over a 20 year-long baseline. Recent X-ray observations via \emph{XMM-Newton} and \emph{NuSTAR}, along with available radio (VLA), mid-IR (\emph{WISE}), near-IR (UKIDSS), optical (SDSS), and UV (\emph{GALEX}) observations, allow for a multi-wavelength analysis to search for evidence of two supermassive black holes. 
\par 
Our paper is organized in the following manner: in Section 2 we analyze new X-ray observation of \quasar{} and evaluate the 0.5--10 keV spectrum for evidence of a binary AGN system; in Section 3 we present the multi-wavlength SED and compare the emission to both single and binary AGN emission models; in Section 4 we discuss our results and test for effects of reddening; and in Section 5 we review our conclusions. We assume a standard $\Lambda$CDM cosmology of $\Omega_{\Lambda}$= 0.7, $\Omega_{M}$ = 0.3, and $H_{0}$ = 70 km s$^{-1}$ Mpc$^{-1}$.

\section{X-ray observations} \label{sec:Xray}
\quasar{} was targeted by \emph{NuSTAR} and \emph{XMM-Newton} in a joint \emph{NuSTAR} Cycle 6 Proposal (PI: Liu, ID: 6061). The quasar was observed for 100 ks and 50 ks with \emph{NuSTAR} and \emph{XMM-Newton} on 2020-08-30 (observation ID 60601009002) and 2020-08-02 UT (observation ID 0870810201). The \emph{NuSTAR} exposure time was set to achieve at least ${\sim}10$ counts under the assumption of a typical optical quasar SED. The \emph{XMM-Newton} exposure time was set to achieve $\sim$200 counts for a $\sim$10\% flux measurement and a simultaneous fit to the X-ray spectral index within $\pm$0.3 and the intrinsic hydrogen column density to an upper limit of ${\sim}10^{21}$ cm$^{-2}$ (both at $\sim$90\% confidence).
\par All errors evaluated in the following section are done at the 95\% confidence level and error bars quoted in the following section are calculated with Monte Carlo Markov Chains via the XSPEC tool {\tt chain}. \\

\subsection{NuSTAR}
We follow the standard process \emph{nupipeline} embedded in the software package NuSTARDAS v1.9.2 to clean the event file of NuSTAR observation. We find no emission consistent with an X-ray point source within 100\arcsec~of the SDSS-listed optical center of \quasar{}. We calculate a 3$\sigma$ upper limit for 3--10 keV and 10--79 keV flux of $8.8\times10^{-14}$ erg s$^{-1}$ and $67\times10^{-14}$ erg s$^{-1}$, respectively, within a circular region with radius of 100\arcsec centered on the optical center. 

\subsection{XMM-Newton}
We clean and process the \emph{XMM-Newton} EPIC pn observation using the \emph{XMM-Newton} Science Analysis System (SAS) software package v18.0.0. The quasar’s net count rate and flux value are determined using XSPEC, version 12.11.1 (\citealt{Arnaud1996}). We generate the event list with the standard pipeline \emph{epproc}. We filtered the event list from high-background time intervals and calculate a good exposure time of 42.59 ks. The quasar was identified as an X-ray point source coincident with the nominal, SDSS-listed optical center of \quasar{}. Counts are extracted from a circular region with radius of 32\arcsec\ centered on the X-ray source center, using a source-free circular region with radius of 90\arcsec\ for the background extraction. The spectrum was been rebinned via the {\tt specgroup} tool to ensure a minimum signal-to-noise ratio of 2 over the 0.3--10 keV band. 
\par To characterize the X-ray emission, and search for evidence of two accreting supermassive black holes, we fit the observed-frame 0.5--10 keV spectrum with 3 different models: Model 1: an absorbed red-shifted power-law ({\tt phabs $\times$ zphabs $\times$ zpow}), and Models 2 and 3:  an absorbed red-shifted broken power-law ({\tt phabs $\times$ zphabs $\times$ zbknpow}, where the photon index values are tied for Model 2). We expect that Model 1 and Model 2 should return results consistent with one-another, however comparing Model 2 and Model 3 allow us to test for the presence of two AGN, which may be contributing X-ray emission from individual accretion disks (and thus, two power-law photon indices may better describe the X-ray emission). 
\par We implement the Cash statistic ({\tt cstat}; \citealt{Cash1979}) to best assess the quality of our model fits.  In particular, we quantify whether Model 3 is preferred over Model 2 by evaluating whether there is a statistically significant improvement in the fit, such that $\Delta C_{\mathrm{stat}}>$ $\Delta N_{\mathrm{fp}} \times$2.71 (where $\Delta N_{\mathrm{fp}}$ represents the difference in number of free parameters between the models; \citealt{Tozzi2006, Brightman2012}), corresponding to a fit improvement with 90\% confidence (\citealt{Brightman2014}).  $K$-corrections are not applied to the \emph{XMM-Newton} data, as we directly measure the flux density from the spectrum. As expected, we find that Model 1 and Model 2 return consistent results; but using the $\Delta C_{\mathrm{stat}}$ criterion stated above, we find that Model 3 results in a significant improvement in the fit compared to Model 2. However, the best-fit values for the power-law photon indices $\Gamma_{1}$ and $\Gamma_{2}$ are unconstrained and pegged to values consistent with the low ($\Gamma_{1} \approx 1$) and high ($\Gamma_{2} \approx 3$) end of the allowed range (where values less than 1 or greater than 3 are usually considered non-physical; see \citealt{Ishibashi2010}). These results may reflect that the X-ray spectrum is consistent with emission from a single AGN, but requires additional components to a simple power-law.
\par
\begin{figure}
\centering
    \includegraphics[width=\linewidth]{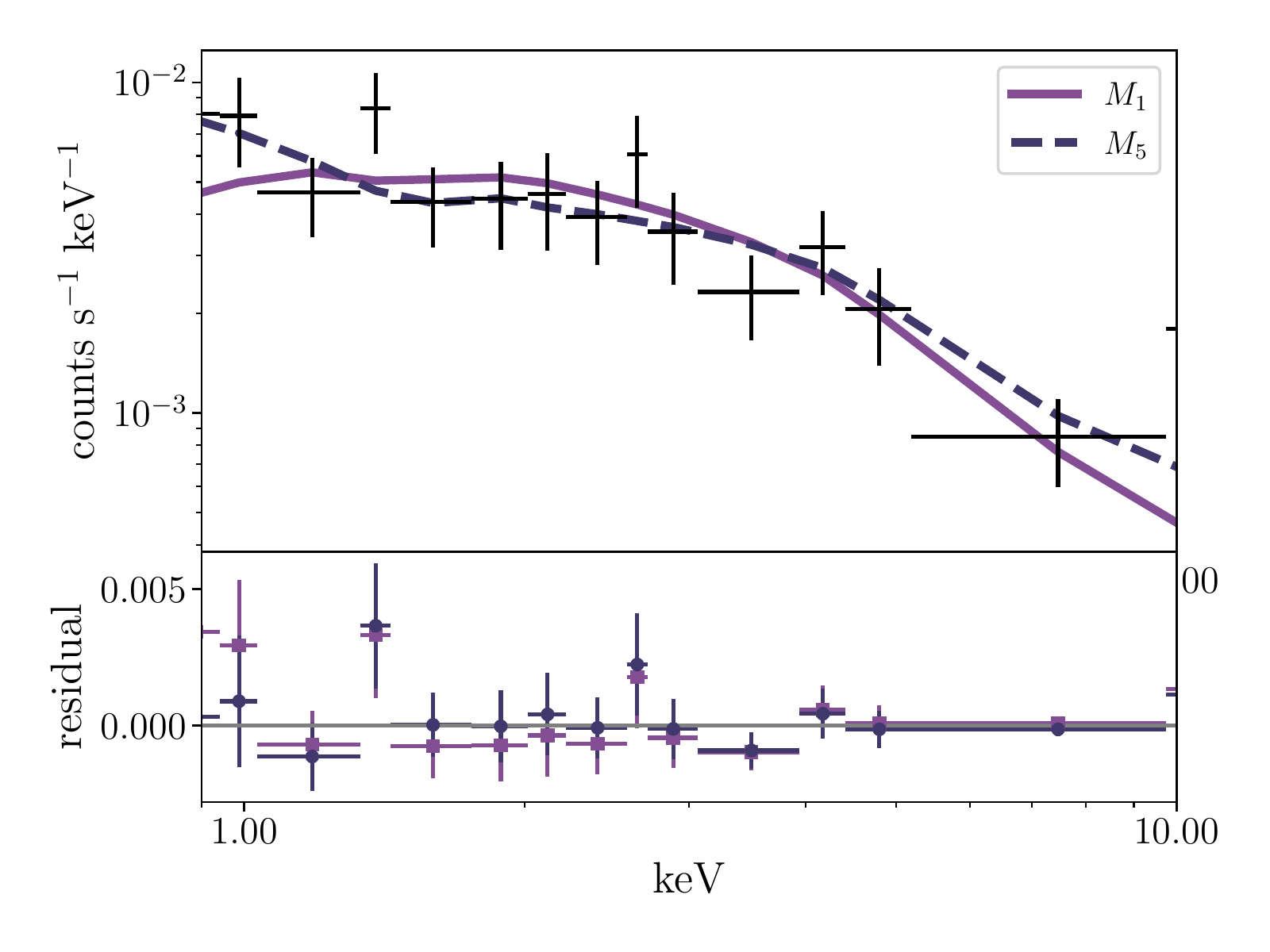}
    \vspace{-3.em}
    \caption{The rest-frame \emph{XMM-Newton} spectrum for \quasar{} (\emph{top}). The spectrum has been rebinned via the {\tt specgroup} tool to ensure a minimum signal-to-noise ratio of 2 over observed-frame 0.3$‑$10 keV band. We show best-fits for both Model 1 (an absorbed red-shifted power-law -- {\tt phabs $\times$ zphabs $\times$ zpow}) and Model 5 (a power-law with diffuse gas -- ({\tt phabs$\times$((zphabs$\times$zpow)+APEC)}). Although Model 5 results in a significantly improved fit compared to Model 1, the posteriors returned by {\tt chain} show that the preferred value for the power-law photon index $\Gamma$ is non-physical. An additional soft emission component may be present within the region of extracted counts, such as extended soft X-rays from hot nuclear gas. Given the non-physical value for $\Gamma$ returned by Model 5, we accordingly assume the simple absorbed power-law fit as our best model (Model 1). We measure a $0.5-10$ keV flux of $9.1_{-3.9}^{+3.1}\times10^{-13}$ erg s$^{-1}$, corresponding to a $2-10$ keV luminosity of $6.1_{-2.0}^{+1.4}\times 10^{44}$ erg s$^{-1}$ at z=1.53.}

\label{fig:xrayspec}
\end{figure}

 Thus, we add several additional components to the absorbed power-law to test if they better describe the X-ray emission. In particular, we look at a partially covered power-law ({\tt phabs$\times$((zphabs$\times$zpow)+zpow)}, where the photon indices of each power-law are tied; Model 4) and a power-law with diffuse gas to account for possible extended soft X-ray emission in the nucleus ({\tt phabs$\times$((zphabs$\times$zpow)+APEC)}, with abundance fixed at solar value; Model 5). We find that the best-fit parameters from Model 4 are consistent with Model 1, where partially absorbed power-law component does not result in a better fit (which is expected, given low-levels of the extragalactic hydrogen column density, $N_{H}$, found across all spectral models; see Table~\ref{tab:spec}). Model 5 results in a significantly improved fit compared to Model 1, meeting our $\Delta C_{\mathrm{stat}}$ criterion. However, similar to Model 3, the posteriors returned by {\tt chain} show that the preferred value for the power-law photon index $\Gamma$ is non-physical and pegged at 1. Evaluating the spectral fits (see Fig.~\ref{fig:xrayspec}), an additional soft emission component may be present within the region of extracted counts, possibly due to extended soft X-rays from hot nuclear gas. However, given the non-physical value for $\Gamma$ returned by Model 5, we accordingly assume the simple absorbed power-law fit as our best model (Model 1). We measure a 0.5--10 keV flux of $9.1_{-3.9}^{+3.1}\times10^{-13}$ erg s$^{-1}$, corresponding to a 2--10 keV luminosity of $6.1_{-2.0}^{+1.4}\times 10^{44}$ erg s$^{-1}$ at $z=1.53$. In Figure~\ref{fig:xrayspec} we show the X-ray spectrum of \quasar{}, along with the fits from Model 1 and Model 5. We list the best-fit values for model parameters in Table~\ref{tab:spec}.

\begin{table}[t]
\begin{center}
\caption{\emph{XMM-Newton} Spectral Fits}
\label{tab:spec}
\begin{tabular*}{\columnwidth
}{ccccccc}
	\hline
	\hline
	\multicolumn{1}{c}{Model} & \multicolumn{1}{c}{$N_{H}$} & \multicolumn{1}{c}{$\Gamma_{1}$} & \multicolumn{1}{c}{$\Gamma_{2}$} & \multicolumn{1}{c}{$kT$} & \multicolumn{1}{c}{$N_{\mathrm{fp}}$} & \multicolumn{1}{c}{$C_{\mathrm{stat}}$} \\
	\multicolumn{1}{c}{(1)} & \multicolumn{1}{c}{(2)} & \multicolumn{1}{c}{(3)} & \multicolumn{1}{c}{(4)} & \multicolumn{1}{c}{(5)} & \multicolumn{1}{c}{(6)} & \multicolumn{1}{c}{(7)}\\
	\hline \\ [-2.5ex]
    1 & $<10^{-2}$ & $1.5_{-0.5}^{+1.0}$ & \dots & \dots & 3 & $41.4$ \\ [0.4ex]
    
    2 & $<10^{-2}$ & $1.5_{-0.5}^{+0.8}$ & $1.5_{-0.5}^{+0.8}$ & \dots & 4 & $41.4$ \\ [0.4ex]
    
    3 & $<10^{-2}$ & $1.1_{-0.1}^{+1.8}$ & $2.8_{-1.7}^{+0.2}$ & \dots & 5 & $37.1$  \\ [0.4ex]
    
    4 & $<10^{-2}$ & $1.5_{-0.4}^{+0.8}$ & $1.5_{-0.4}^{+0.8}$ & \dots & 4 & $41.4$  \\ [0.4ex]
    
    5 & $<10^{-2}$ & $1_{-0}^{+1.9}$ & \dots & $1.2_{-0.5}^{+8.8}$ & 5 & $35.6$ \\ [0.4ex]
	\hline 
	\hline
\end{tabular*}

\end{center}
Note. -- Columns: (1) Model number, where Model 1 is an absorbed red-shifted power-law ({\tt phabs $\times$ zphabs $\times$ zpow}), Model 2 and 3 are an absorbed red-shifted broken power-law ({\tt phabs $\times$ zphabs $\times$ zbknpow}, where the photon index values are tied for Model 2), Model 4 is a partially covered power-law ({\tt phabs$\times$(zphabs$\times$zpow)+zpow)}), and Model 5 is a power-law with diffuse gas ({\tt phabs$\times$(zphabs$\times$zpow)+APEC)}; (2) the best-fit extragalactic column density in units of 10$^{22}$ cm$^{-2}$; (3) the best-fit spectral index for $\Gamma_{1}$; (4) the best-fit spectral index for $\Gamma_{2}$ (only relevant for Model 3); (5)  best-fit spectral value for $kT$ (only relevant for Model 5); (6) the number of free parameters for a given model; (7) the Cash statistic for the best-fit. Error bars represent the 95\% confidence level of each distribution.
\end{table}

\section{Multi-wavelength SED} \label{sec:SED}
In the following section, we construct a multi-wavelength SED of \quasar{} (see also \citealt{Liao2021}, where this data set is first presented). Similar to the analysis presented in \cite{Foord2017b}, we combine all available multi-wavelength observations, and compare the SED to the standard non-blazar AGN SEDs presented in \cite{Shang2011}. To correct for the effective narrowing of the filter width with redshift, we adopt the $K$-correction relation as presented in \cite{Richards2006}. In particular, the $K(z)$ relation for power-law continuum is given by $K = -2.5 (1 + \alpha_{\nu})\log{(1 + z)}$, assuming $F_{\nu} \propto \nu^{-\alpha_{\nu}}$.
\par The available SED observations include a radio flux measurement from the Karl G. Jansky Very Large Array (VLA; \citealt{Thompson1980}), where the rest-frame 6 GHz luminosity is measured in \cite{Chen2021}; archival MIR photometry from the Wide-field Infrared Survey Explorer (\emph{WISE}, \citealt{Wright2010}); archival NIR photometry from the UKIRT Infrared Deep Sky Survey (UKIDSS, \citealt{Lawrence2007}); archival optical photometry from the Sloan Digital Sky Survey (SDSS, \citealt{York2000}); and archival UV photometry from the Galaxy Evolution Explorer (\emph{GALEX}, \citealt{Martin2005}). We correct the archival MIR--UV magnitudes for extinction using dust maps from \cite{Schlafly&Fink2011} and reddening curves from \cite{Fitzpatrick99}. $K$-corrections are then applied assuming values of $\alpha = -1.0, -0.5, -1.57,$ for the IR, optical, and UV measurements (\citealt{Shang2011, Ivezic2001, Richards2006}). We also include our X-ray observations from \emph{XMM-Newton} \citep{Jansen2001}, and \emph{NuSTAR} \citep{Harrison2013}. The luminosities for each rest-frame frequency are listed in Table~\ref{tab:mw}.
\par
\begin{figure*}
\centering
    \includegraphics[width=14cm]{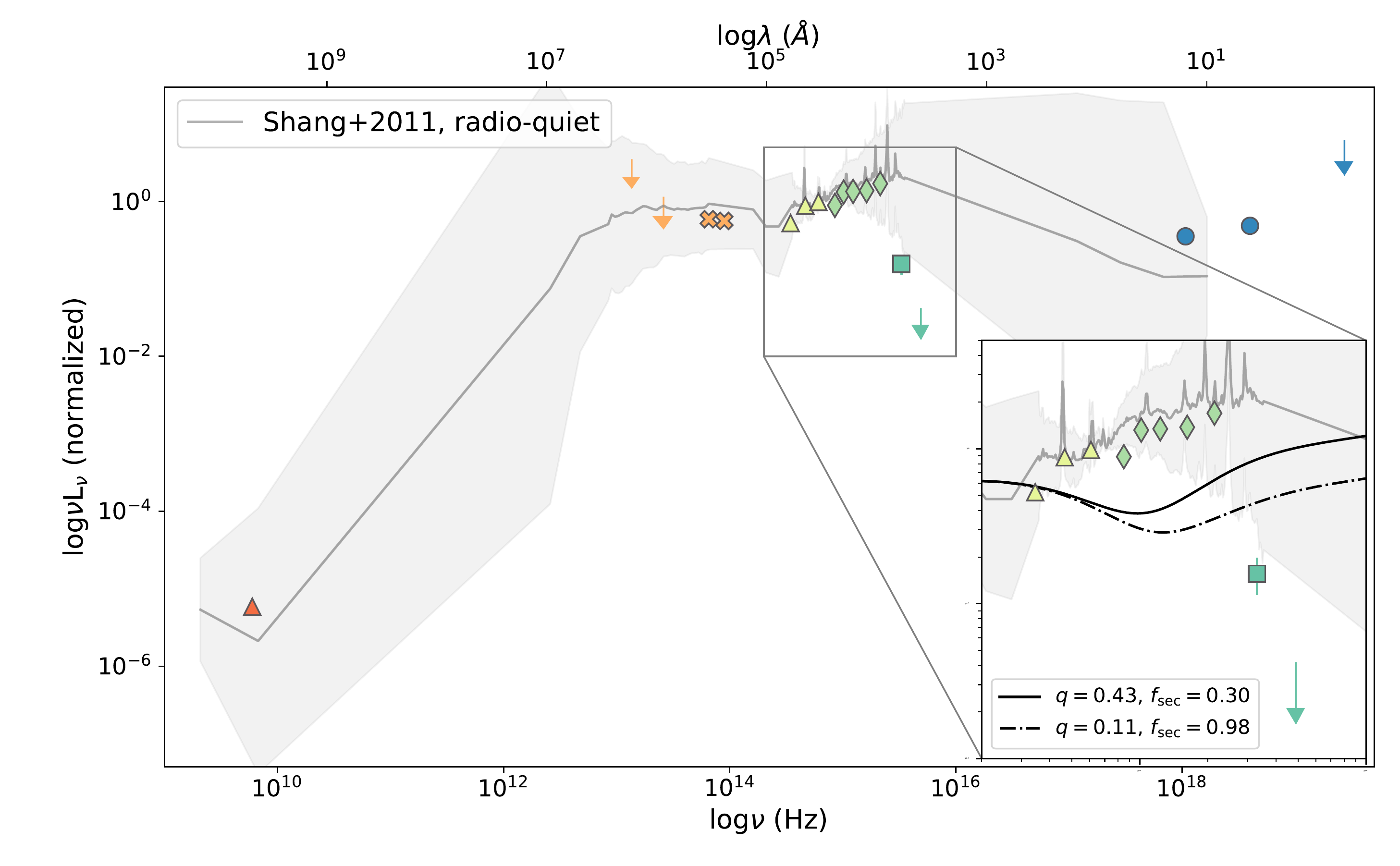}
    \caption{The rest-frame multi-wavelength spectral energy density of \quasar{}. We plot the radio flux density upper limit from VLASS (red pentagon), MIR photometry from \emph{WISE} (orange ``x" markers for the W2 and W1 filter detections, and upper-limits for the W3 and W4 filters), NIR photometry from UKIDSS (yellow triangles), optical photometry from SDSS (green diamonds), UV photometry from \emph{GALEX} (dark green square for the NUV detection, and upper limit for the FUV), and X-ray photometry from \emph{XMM-Newton} (dark blue circles) and \emph{NuSTAR} (upper limit). In gray we overplot the composite non-blazar radio-quiet quasar SED from \cite{Shang2011}. In general we find a good agreement between the SED of \quasar{} and that of the composite quasar SED. However, \quasar{} appears to have a deficit of emission between the NIR and FUV bands, significantly dropping off near $\sim$1400~${\rm\AA}$. In the inset we show various models for a possible ``notch" in the accretion disk of a binary AGN, with different mass ratios ($q$) and accretion rates for the primary, $f_{pri}$, and secondary, $f_{sec}$ (where $f_{pri}+f_{sec}=1$, see Section~\ref{subsec:binary} for more detail). Although lower values of $q$ and higher values $f_{sec}$ will result in deeper and wider notch, we find that the most extreme values fail to accurately capture the FUV \emph{GALEX} upper-limit.}
\label{fig:MW_SED}
\end{figure*}

 In Figure~\ref{fig:MW_SED} we plot the full multi-wavelength SED of \quasar{} where, following the normalization procedure in \citealt{Shang2011}, the flux density of our data is normalized to $\lambda\approx4200$ ${\rm\AA}$. This value is estimated by interpolating between our bluest UKIDSS data point, at rest-frame $\lambda\approx4934$~${\rm\AA}$, and our reddest SDSS data point, at rest-frame $\lambda\approx3530$~${\rm\AA}$. A simple comparison between the data and the \cite{Shang2011} SED shows a good agreement between the emission of \quasar{} and that of a radio-quiet AGN. However, there is a clear deficit of emission from \quasar{} between the NIR and UV frequencies, with a large drop in emission near 1400 ${\rm\AA}$. Furthermore, given the FUV \emph{GALEX} upper-limit at rest-frame $\sim600$ ${\rm\AA}$, it is possible that the slope of the drop is larger than presented in Figure~\ref{fig:MW_SED}. 

\begin{table}[t]
\begin{center}
    \caption{Multi-wavelength Luminosity Values}
\label{tab:mw}
\begin{tabular*}{\columnwidth
}{ccccccc}
	\hline
	\hline
	\multicolumn{1}{c}{Filter/Detector} & \multicolumn{1}{c}{Telescope/Survey} & \multicolumn{1}{c}{$\log{\nu}$} & \multicolumn{1}{c}{$\log{\nu L_{\nu}}$} \\
	\multicolumn{1}{c}{(1)} & \multicolumn{1}{c}{(2)} & \multicolumn{1}{c}{(3)} & \multicolumn{1}{c}{(4)}\\
	\hline \\ [-2.5ex]
    C-band & VLA & $9.78$ & $39.98\pm0.63$\\ [0.4ex]
    
    W4 & \emph{WISE} & $13.5$ & $<45.76$  \\ [0.4ex]
    W3 & \emph{WISE} & $13.8$ & $<45.28$  \\ [0.4ex]
    W2 & \emph{WISE} & $14.2$ & $44.99\pm0.11$  \\ [0.4ex]
    W1 & \emph{WISE} & $14.4$ & $44.96\pm0.24$  \\ [0.4ex]
    
    J & UKIDSS & $14.5$ & $45.21\pm0.05$ \\ [0.4ex]
    K & UKIDSS & $14.7$ & $45.16\pm0.06$ \\ [0.4ex]
    H & UKIDSS & $14.8$ & $44.94\pm0.07$ \\ [0.4ex]
    
    $z$ & SDSS & $14.93$ & $45.17\pm0.14$   \\ [0.4ex]
    $i$ & SDSS & $15.00$ & $45.34\pm0.39$   \\ [0.4ex]
    $r$ & SDSS & $15.01$& $45.35\pm0.23$   \\ [0.4ex]
    $g$ & SDSS & $15.21$ & $45.36\pm0.24$   \\ [0.4ex]
    $u$ & SDSS & $15.32$ & $45.45\pm0.31$   \\ [0.4ex]
    
    NUV & $\emph{GALEX} $ & $15.52$ & $44.12\pm0.45$ \\ [0.4ex]
    FUV & $\emph{GALEX} $ & $15.69$ & $<43.50$ \\ [0.4ex]
    
    EPIC pn & $\emph{XMM-Newton} $ & $18.16$ & $44.79^{+0.55}_{-0.50}$ \\ [0.4ex]
    \dots  & $\emph{NuSTAR} $ & $19.04$ & $<46.02$ \\ [0.4ex]
	\hline 
	\hline
\end{tabular*}
\end{center}
Note. -- Columns: (1) Filter or detector; (2) Telescope or survey; (3) rest-frame frequency assuming a redshift of $z=1.53$, in units of hertz. The \emph{XMM-Newton} and \emph{NuSTAR} frequency corresponds to the central rest-frame frequency of their respective observing ranges (i.e., 6 and 45 keV); (4) extinction- and $K$-corrected luminosity assuming a luminosity distance $D_{L}=11.36$ Gpc, in units of erg s$^{-1}$. Please see Section 3 for details on extinction values and $K$-corrections applied.
\end{table}
\vspace{-0.1cm}
\subsection{Comparison to binary AGN accretion models}
\label{subsec:binary}
\par This drop in emission in the SED may be a result a binary AGN accretion mode. In particular, if \quasar{} is a binary AGN system with separation a = 200 $R_{S}$ (where $R_{S} = 2GMc^{-2}$ is the Schwarzchild radius for a black hole with mass $M$), as estimated in \cite{Liao2021}, the binary is well into the gravitational-wave dominated regime. Here, circumbinary
accretion is likely and we may expect that individual accretion disks have formed around each SMBH. In such a scenario, any radiation that a standard accretion disk would radiate between the inner edge of the circumbinary disk and the tidal truncation radii of the minidisks will
be missing. This missing emission can produce a notch in the thermal
continuum spectrum (e.g., \citealt{Gultekin&Miller2012, Kocevski2012, Roedig2012, Tanaka2012, Tanaka&Haiman2013, Roedig2014}, and see \citealt{Farris2015b} for simulations where notches become obscured). 
\par Following a similar procedure as outlined in \cite{Foord2017b}, we use the analytical calculations derived in \cite{Roedig2014} to analyze how the SED shape is affected by the presence of a notch. In particular, we model the specific luminosity integrated from the circumbinary disk and the two minidisks assuming that the primary and secondary BHs are accreting at rates $\dot{M}_{pri}=f_{pri}\dot{M}$ and $\dot{M}_{sec}=f_{sec}\dot{M}$. Here, $\dot{M}_{pri}$ and $\dot{M}_{sec}$ are the mass accretion rates of the primary and secondary, and $\dot{M}$ is the circumbinary accretion rate. These calculations assume that the circumbinary disk is in inflow equilibrium, such that $f_{pri}+f_{sec}=1$.  Values of $f_{pri}$ and $f_{sec}$ depend on the mass ratio of the system, $q$, and simulations have shown that for most values of mass ratio, $f_{sec} > f_{pri}$ (but as $q$ increases towards equal mass, the accretion rate ratio goes to unity, see \citealt{Farris2014}). 
\par Hydrodynamical simulations of circumbinary accretion disks around binary SMBHs with a range of mass ratios ($0.026<q<1.0$) have shown that significant periodicity in the accretion rates occurs only at $q>0.1$ (\citealt{Farris2014}). At these mass ratios, the binary
torques are strong enough to lead to periodicity in the accretion rates, and thus the light curve of the system. Previously, the light curves of \quasar{} were modeled with both $q=0.11$ and $q=0.43$ (two mass ratio regimes probed by \citealt{Farris2014}); however, the fits to the data did not strongly prefer one model over the other, and a value of $q=0.11$ was adopted to interpret the results (\citealt{Liao2021}).
\par In Figure~\ref{fig:MW_SED} we show two examples of SED models with notches for both $q=0.11$ and $q=0.43$. Here, we adopt the predicted $f_{pri}$ and $f_{sec}$ values expected for each mass ratio as estimated in \cite{Farris2014}. As evident, as the mass ratio decreases and the secondary's accretion rate dominates, the notch occurs at shorter wavelengths and deepens; this is a result of the primary SMBH’s accretion flow barely contributing to the total SED (\citealt{Roedig2014}). Because we don't expect significant modulation in the accretion rates at $q<0.1$, our mass ratio model of $q=0.11$ represents the lowest and deepest notch expected for this system. We find that this model is unable to match the shape of \quasar{}'s SED, and in particular predicts a lower level of emission between rest-frame 1400$-$3500 ${\rm\AA}$~(SDSS) and a higher level of emission at $\lambda<1400$~${\rm\AA}$~(\emph{GALEX}) than observed. \par Lastly, disregarding the mass ratio constraint of $q>0.1$ for significant accretion modulation, extremely low (high) values of $q$ ($f_{sec}$) can further deepen the notch to potentially match the \emph{GALEX} data points. However the size of the notch will also widen -- resulting in larger differences between the predicted and observed rest-frame optical and NUV emission (UKIDSS and SDSS observations, see Figure~\ref{fig:MW_SED}).
\vspace{1cm}
\subsubsection{Enhanced Hard X-ray Emission}
\vspace{-0.1cm}
\par Many computational results on binary mergers indicate that binary SMBHs should have enhanced hard X-ray emission relative to a single SMBH (\citealt{Roedig2014, Farris2015b, Ryan2017, Tang2018, dAscoli2018}), a result of supersonic streams from the circumbinary disk shocking at the minidisk edges. Yet, results from these simulations have a wide range in the predicted energy at which these enhancements should occur, from tens of keV up to over 100 keV. Although \cite{Roedig2014} predict a Wien-like spectrum may adequately describe the emission, the large range of possible peak temperatures does not allow us to carry out an in-depth analysis of a possible hot-spot X-ray contribution to the multi-wavelength SED. 
\par However, using the derivations presented in \cite{Roedig2014}, we can estimate an approximate peak temperature of an additional Wien-like spectrum by calculating the post-shock temperature, $T_{ps}$, which is proportional to the binary’s semimajor axis, $a$, and mass ratio, $q$: $T_{ps} \propto (a/50R_{S})^{-1}\times(1+q^{0.7})^{-1}$ (see \citealt{Roedig2014} for more details). We consider possible temperatures for the stream-shocking emission from \quasar{}, given an assumed semimajor axis $a=200~R_{S}$ and a range of $q$ values between 0.1$-$1.0. Adopting the assumption presented in \citealt{Roedig2014} that a mass ratio $q=1.0$ and semimajor axis $a=50~R_{S}$ will result in an additional Wien-like spectrum with peak energy 100 keV, we estimate that excess emission from hot spots in \quasar{} could peak between rest-frame 25$-$42 keV, or observed-frame 10$–$16 keV. Although our \emph{XMM-Newton} falls below this energy window, the spectrum shows no evidence of excess hard X-ray emission with respect to an absorbed power-law model. Furthermore, our \emph{NuSTAR} upper-limit does not allow us to measure for the presence of enhanced hard X-ray emission at higher energies. Additional studies on the hard X-ray emission associated with binary AGN (especially in the unequal-mass regime), will allow for more rigorous hard X-ray analyses on binary AGN candidates in the future.
\par Lastly, we note that the enhanced X-ray emission expected from binary AGN raises the possibility of binaries having different X-ray spectral indices (see Section 2 for our analysis on the X-ray spectrum) or optical to X-ray spectral indices ($\alpha_{\rm ox}$, see, e.g., \citealt{Yuan1999, Vignali2003, Strateva2005, Steffen2006, Just2007, Kelly2008, Lusso2012}) than single AGN. This ratio is defined as $\alpha _{\rm ox} = -\log{[L_{\rm 2~keV}/L_{2500}]}/2.605$, and has been measured to strongly correlate with the optical luminosity of the AGN at 2500 ${\rm\AA}$ (e.g., \citealt{Silverman2005, Lusso2012}). We estimate the approximate rest-frame 2500 ${\rm\AA}$ flux density using the available SDSS $r$ band photometry (which corresponds to rest-frame emission at $\approx$ 2440 \AA), and we use the \emph{XMM-Newton} fit results from Model 1 (see Section 2) to estimate the 2 keV flux density. We calculate an $\alpha_{\rm ox}\approx1.6$, consistent within the error of the range expected from single AGN with similar 2500 ${\rm\AA}$ luminosities (see \citealt{Lusso2012}). 
\begin{figure}
\centering
    \includegraphics[width=8cm]{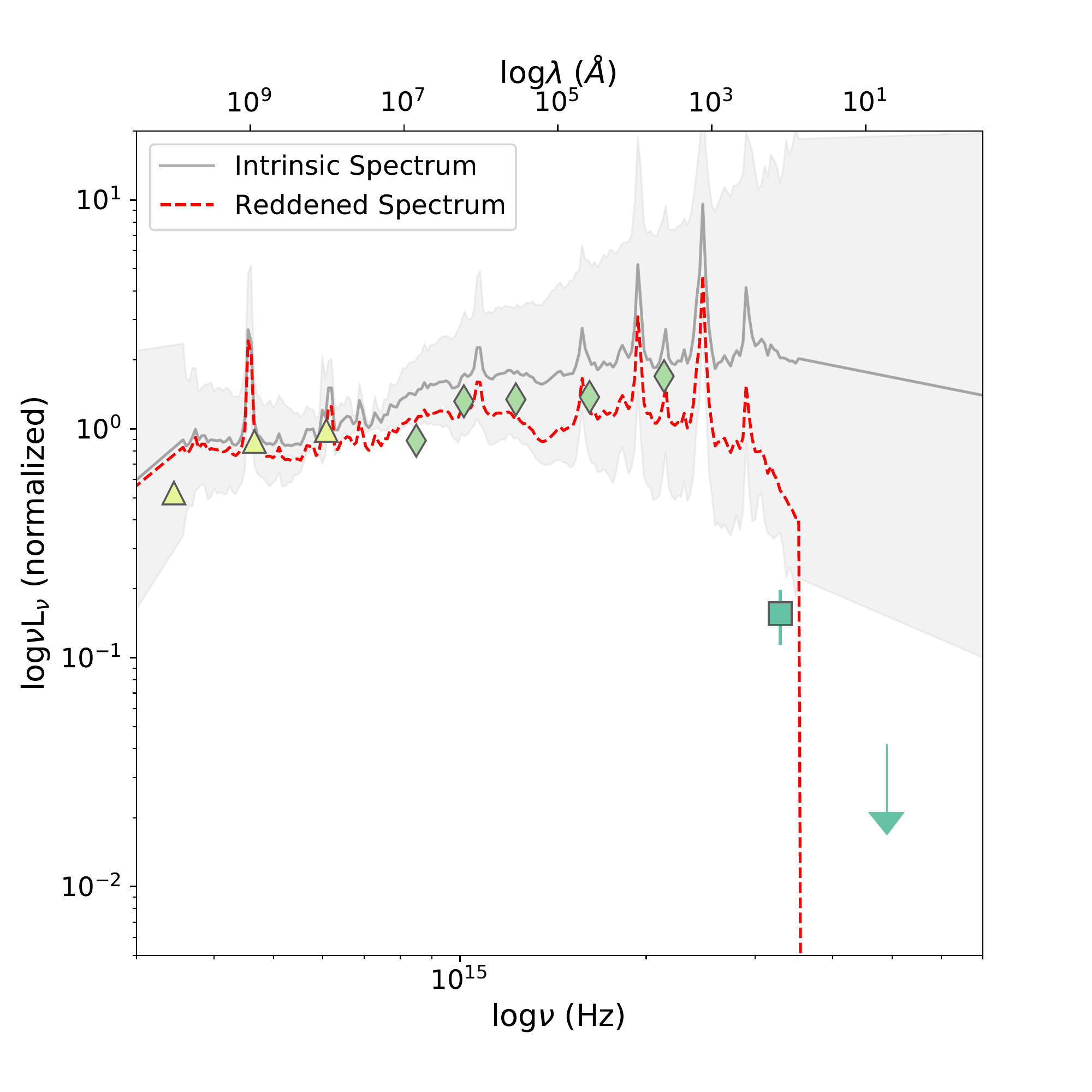}
    \caption{The best-fit reddened spectrum (red dashed line) when fitting the composite radio-quiet SED from \cite{Shang2011} (gray line) to the photometric data points of \quasar{} (between rest-frame 87300~${\rm\AA}$~and 610~${\rm\AA}$). We plot NIR photometry from UKIDSS (yellow triangles), optical photometry from SDSS (green diamonds), and UV photometry from \emph{GALEX} (dark green square for the NUV detection, and upper limit for the FUV). We apply the reddening law from \cite{Fitzpatrick99}, although we find consistent results when using reddening law presented in \cite{Goobar2008} (which has been used in past analyses of binary AGN candidate Mrk 231). We find best-fit values of $A_{V}=0.17$ and $R(V)=2.54$. Overall, once accounting for circumstellar reddening, the composite single quasar spectrum agrees well the measured emission of \quasar{}.}
\label{fig:red}
\end{figure}
\section{Possible effects of reddening}
In the following section we discuss alternative physical explanations for the observed multi-wavelength SED of \quasar{}, which is not well explained by either a standard AGN or a binary AGN system. In particular, we investigate whether the effects of reddening can match the steep drop off seen near $\sim$1400 ${\rm\AA}$. 
\par The unusual SED of \quasar{}, with infrared and optical emission typical of an AGN but strongly cut off through the near UV, is similar (albeit less extreme) to the SED of contentious binary AGN candidate Mrk 231 (see \citealt{Smith1995, Veilleux2013, Leighly2014, Yan2015, Leighly2016}). The binary AGN hypothesis proposed for Mrk 231 is that the smaller-mass black hole accretes as a thin disk, dominating the weak UV emission, while the larger-mass black hole radiates inefficiently as an Advection Dominated Accretion Flow (ADAF), and both are surrounded by a circumbinary disk (dominating the optical and IR emisison). Here, emission blueward of the near UV is dominated by accretion onto the smaller mass black hole and thus the SED is expected to have a steep drop-off toward the UV, near the inner edge of the circumbinary disk (\citealt{Yan2015}). However, it has been shown that the unusual shape is consistent with circumstellar reddening (\citealt{Leighly2014}), and that if the observed FUV continuum is intrinsic, it fails by a factor of 100 in powering the observed strength of the near-infrared emission lines (\citealt{Leighly2016}).
\par Following the logic presented in \cite{Leighly2014}, we apply a reddening correction from \cite{Fitzpatrick99} to the composite radio-quiet quasar SED from \cite{Shang2011} to fit the photometric data points of \quasar{} between rest-frame 87300~${\rm\AA}$~and 610~${\rm\AA}$~(i.e., the \emph{WISE}, UKIDSS, SDSS, and \emph{GALEX} data points). We fit for the best values of $A_{V}$ and $R(V)$ via the {\tt python} non-linear least-squares fitting package {\tt lmfit}. In addition to the reddening relation presented in \cite{Fitzpatrick99}, we also use the analytical correction presented in \cite{Goobar2008}, which assumes a spherical scattering medium (and used by \citealt{Leighly2014} when analyzing Mrk 231). Although this reddening law is typically used to describe low values of $R(V)$ around Type Ia supernovae, a spherical geometry with significant
optical depth (and multiple scatterings) allows for the removal of more
blue photons from the line of sight than a screen. 
\par The best-fit results of both reddening laws are consistent within the errors, although the reddening correction from \cite{Fitzpatrick99} yields a slightly better fit (as determined from the Bayesian information criterion values). We find best-fit values of $A_{V}=0.17$ and $R(V)=2.54$. In Figure~\ref{fig:red} we plot the results of our best-fit reddened quasar SED along with the photometric data points from the SED of \quasar{}. With the addition of circumstellar reddening, the composite quasar spectrum agrees well with the measured emission of \quasar{}. The best-fit $R(V)$ value is indicative of dust similar to that in the Milky Way (where $R(V)\sim3.1$).
\par Although we find the SED of \quasar{} best described by a single AGN obscured by circumstellar dust, it is possible that the system is a binary where a notch has been masked by (i) circumstellar dust, (ii) the tail-end of a Wien-like spectrum that peaks at hard ($>$10 keV) energies, or (iii) a combination of both. Follow-up spectroscopy in the IR, which is relatively free of the effects of reddening, can help disentangle whether circumstellar reddening is acting on its own. For example, \cite{Leighly2016} analyze observed He \RN{1} ($\lambda=10830$ ${\rm\AA}$), P$\beta$ ($\lambda=12818$ ${\rm\AA}$), P$\alpha$ ($\lambda=18751$ ${\rm\AA}$), and C IV ($\lambda=1549$ ${\rm\AA}$) emission lines of Mrk 231 with {\it Cloudy} to show that the measured equivalent widths and emission-line ratios in the IR are not reproducible with the binary AGN model proposed by \cite{Yan2015}. 

\section{Conclusions}
In this work we present a multi-wavelength analysis of binary AGN candidate \quasar{}. \quasar{} was part of a large systematic search for long-term periodic light curves in 625 quasars, and was flagged as a binary AGN candidate based on a significant periodicity measured in 20 year SDSS--DES data (\citealt{Chen2020, Liao2021}). New X-ray observations via \emph{XMM-Newton} and \emph{NuSTAR}, along with available radio (VLA), mid-IR (\emph{WISE}), near-IR (UKIDSS), optical (SDSS), and UV (\emph{GALEX}) observations, have been combined to search for evidence of two accreting supermassive black holes. The main results and implications of this work can be summarized as follows.
\begin{enumerate}
    \item We analyze new X-ray observations of \quasar{} (obtained by \emph{NuSTAR} and \emph{XMM-Newton}). We find no emission consistent with an X-ray point source within 100\arcsec~of the SDSS-listed optical center of \quasar{} when analyzing the \emph{NuSTAR} data. We calculate a 3$\sigma$ upper limit for 3--10 keV and 10--79 keV flux of $8.8\times10^{-14}$ erg s$^{-1}$ and $67\times10^{-14}$ erg s$^{-1}$, respectively. We identified the quasar as an X-ray point source in the \emph{XMM-Newton}, and search for evidence of two accreting supermassive black holes by fitting the observed-frame 0.5$-$10 keV spectrum with 5 different models. Although an additional soft emission component may be present within the region of extracted counts, a simple absorbed power-law remains the best-fit model, as expected from a single AGN. We measure a 0.5--10 keV flux of $9.1_{-3.9}^{+3.1}\times10^{-13}$ erg s$^{-1}$, corresponding to a $2-10$ keV luminosity of $6.1_{-2.0}^{+1.4}\times 10^{44}$ erg s$^{-1}$ at $z=1.53$.
    \item We combine all available multi-wavelength observations and compare the SED to a standard non-blazar AGN SED. The available SED observations include a radio flux measurement from the VLA, MIR photometry from \emph{WISE}, NIR photometry from UKIDSS, optical photometry from SDSS, and UV photometry from \emph{GALEX}. We find a good agreement between the SED of \quasar{} and that of a standard AGN; however there is a clear deficit of emission from \quasar{} between the optical and UV frequencies, with a large drop in emission near rest-frame 1400 ${\rm\AA}$.
    \vspace{-0.1cm}
    \item We investigate whether the drop in emission in the SED may be a result of circumbinary accretion, where individual disks have formed around each SMBH. Using analytical calculations derived in \cite{Roedig2014}, we analyze how the SED of a standard quasar could be affected by the presence of a notch. However, even at the most extreme values of mass ratios and accretion rates, we find the model is unable to match the shape of \quasar{}'s SED. 
    \item We estimate an approximate peak temperature of an additional Wien-like spectrum in the hard X-rays, as a result of possible stream shocking in a binary AGN system. We find that excess emission from hot spots in \quasar{} could peak between rest-frame 25$-$42 keV, or observed-frame 10$-$16 keV. Although our \emph{XMM-Newton} falls below this energy window, the spectrum shows no evidence of excess hard X-ray emission with respect to an absorbed power-law model. Furthermore, our \emph{NuSTAR} upper-limit does not allow us to measure for the presence of enhanced hard X-ray emission at higher energies. We search for other evidence of enhanced X-ray emission by analyzing $\alpha_{\rm OX}$, and calculate a value of 1.6, consistent within the error of the range expected from a single AGN.
    \item We investigate whether the effects of circumstellar reddening can match the steep drop-off in the multi-wavelength SED near 1400 ${\rm\AA}$. Following the logic presented in \cite{Leighly2014}, we apply a reddening correction from \cite{Fitzpatrick99} to a standard non-blazar AGN SED to fit the photometric data points of \quasar{} between rest-frame 87300 ${\rm\AA}$~and 610 ${\rm\AA}$. With the addition of circumstellar reddening, the standard quasar spectrum agrees well with the measured emission of \quasar{}. We find best-fit values of $A_{V}=0.17$ and $R(V)=2.54$.
\end{enumerate}
We have shown through various analyses that there is an absence of evidence supporting \quasar{} as a binary AGN system. However, given the small number of currently confirmed binary AGN, the best method to distinguish a binary AGN from a single AGN is consistenly changing. These studies are further complicated by the fact that analyses searching for signs of circumbinary accretion will likely be dependent on the unique parameters of a given binary AGN system. For \quasar{}, future observations of IR emission lines can be used to better understand whether a binary AGN accretion model is able to power the emission seen at longer wavelengths. Overall, hard X-ray emission signatures may be the most telling sign of circumbinary accretion\,---\,however, current results from simulations are extremely model-dependent (with a wide range of predicted energies where enhancements should occur). A more rigorous analysis of the binary AGN hypothesis for \quasar{} can be made with future studies on the expected hard X-ray emission associated with binary AGN (especially in the unequal-mass regime). 

\vspace{1 cm}
AF acknowledges support by the Porat Postdoctoral
Fellowship at Stanford University. XL and YCC acknowledge support by NSF grant AST-2108162 and NASA grant 80NSSC21K0060. YCC acknowledges support by the government scholarship to study aboard from the ministry of education of Taiwan and the Illinois Survey Science Graduate Student Fellowship.
\emph{NuSTAR} is a project led by the California Institute of Technology (Caltech), managed by the Jet Propulsion Laboratory (JPL), and funded by the National Aeronautics and Space Administration (NASA). We thank the \emph{NuSTAR} Operations, Software and Calibrations teams for support with these observations. 
This research has made use of the \emph{NuSTAR}  Data Analysis Software (NuSTARDAS) jointly developed by the ASI Science Data Center (ASDC, Italy) and the California Institute of Technology (Caltech, USA).
This work is also based on observations obtained with \emph{XMM-Newton}, an ESA science mission with instruments and contributions directly funded by ESA Member States and NASA. This work makes use of data obtained as part of the Karl G. Jansky Very Large Array (VLA), UKIRT Infrared Deep Sky Survey (UKIDSS), Wide-field Infrared Survey (\emph{WISE}), Sloan Digital Sky Survey (SDSS), and the Galaxy Evolution Explorer (\emph{GALEX}). The National Radio Astronomy Observatory is a facility of the National Science Foundation operated under cooperative agreement by Associated Universities, Inc. UKIDSS and \emph{WISE} are joint projects of the University of California, Los Angeles, and the Jet Propulsion Laboratory/California Institute of Technology, funded by the National Aeronautics and Space Administration. Funding for the Sloan Digital Sky Survey has been provided by the Alfred P. Sloan Foundation, the U.S. Department of Energy Office of Science, and the Participating Institutions. \emph{GALEX} is a NASA Small Explorer, launched in 2003 April. We acknowledge NASA’s support for construction, operation, and science analysis for the \emph{GALEX} mission, developed in cooperation with the Centre National d’Etudes Spatiales of France and the Korean Ministry of Science and Technology. Lastly, this research has made use of NASA’s Astrophysics Data System.

%

\vspace{5mm}
\facilities{\emph{NuSTAR} - The NuSTAR (Nuclear Spectroscopic Telescope Array) mission, \emph{XMM-Newton} X-Ray Multimirror Mission satellite, VLA, \emph{WISE}, UKIDSS, SDSS, \emph{GALEX}}


\software{NuSTARDAS,
          SAS,  
          XSPEC, 
          }



\bibliography{foord.bib}
\bibliographystyle{aasjournal}



\end{document}